\documentclass[traditabstract]{aa}
\usepackage[utf8]{inputenc}
\usepackage{amsmath}
\usepackage{amssymb}
\usepackage{microtype}
\usepackage{natbib}
\usepackage{graphicx}
\usepackage{txfonts}
\usepackage{color}
\usepackage{multirow}
\usepackage{xspace}
\usepackage{ulem}
\bibpunct{(}{)}{;}{a}{}{,} 
\newcommand{\src}{\object{4U~0142$+$61}\xspace}

\begin{document}

\title{An observational argument against accretion in magnetars}
\author{V.\,Doroshenko\inst{1,2} \and A.\, Santangelo\inst{1} \and V. F.\,Suleimanov\inst{1,2,3} \and S. S.~Tsygankov\inst{4,2}}
\institute{Institut für Astronomie und Astrophysik, Sand 1, 72076 Tübingen, Germany
\and
Space Research Institute of the Russian Academy of Sciences, Profsoyuznaya Str. 84/32, Moscow 117997, Russia
\and 
Astronomy Department, Kazan (Volga region) Federal University, Kremlyovskaya str. 18, 420008 Kazan, Russia
\and 
Department of Physics and Astronomy,  FI-20014 University of Turku, Finland}

\abstract{The phenomenology of anomalous X-ray pulsars is usually interpreted
within the paradigm of very highly magnetized neutron stars, also known as
magnetars. According to this paradigm, the persistent emission of anomalous
X-ray pulsars (AXPs) is powered by the decay of the magnetic field. However, an
alternative scenario in which the persistent emission is explained through
accretion is also discussed in literature. In particular, AXP~\src has been
suggested to be either an accreting neutron star or a white dwarf. Here, we rule
out this scenario based on the the observed X-ray variability properties of the
source. We directly compare the observed power spectra of \src\  {and of
two other magnetars, 1RXS~J170849.0$-$400910 and 1E~1841$-$045} with that of
the X-ray pulsar 1A~0535+262, and of the intermediate polar GK~Persei.
{In addition, we include a bright young radio pulsar PSR B1509-58 for
comparison.} We show that, unlike accreting sources, no aperiodic variability
within the expected frequency range is observed in the power density spectrum
of {the magnetars and the radio pulsar}. Considering that strong
variability is an established feature of all accreting systems from young
stellar objects to super-massive black holes {and the absence of the
variability reports from other magnetars, we conclude that our results
also indicate}  that magnetars, in general, are not powered by accretion.}

\authorrunning{V. Doroshenko et al.}
\maketitle

\section{Introduction} 
{Magnetars, including soft gamma ray repeaters (SGRs) and anomalous
X-ray pulsars (AXPs), are defined as isolated neutron stars (NSs) emitting from
radio to X-rays, and they are presumably powered by the dissipation of their very strong
magnetic fields ($\sim10^{14}$\,G,
\citealt{2002ApJ...574..332T,2008A&ARv..15..225M})}. {Such fields are
inferred from their rapid spin-down and extreme flaring activity
\citep{1982ApJ...260..371K,1992ApJ...392L...9D,2002ApJ...574..332T}.}
{While fields of a similar magnitude have also been suggested for some
accreting pulsars
\citep{2010A&A...515A..10D,2012MNRAS.425..595R,2016MNRAS.457.1101T}, the
emission in this case is clearly powered by accretion rather than field decay,
so these objects are generally not considered magnetars.}

{On the other hand,} accretion from a fall-back fossil disk, surviving
the supernova explosion onto a magnetized NS
\citep{1995A&A...299L..41V,2001ApJ...554.1245A}, or an isolated white dwarf
(WD) \citep{2014PASJ...66...14C,2020arXiv200407963B} has also been invoked to
explain the persistent X-ray emission and spin evolution of AXPs. A serious
flaw in {this scenario} is that it cannot explain the giant flares
observed from SGRs, which must still be powered by the decay or reconnection
of ultra-strong magnetic fields close to the surface of the NS and triggered by
crustal shifts. {Accretion is thus currently not considered as a
mainstream explanation for the AXP phenomenon.}

{Nevertheless, significant observational efforts aimed at the detection of
optical or infrared emission from cool fossil disks, potentially powering
accretion, have been undertaken and indeed were successful in revealing the
presence of a disk
\citep{2000Natur.408..689H,2001ApJ...556..399K,2006ApJ...649L..87E,2006Natur.440..772W,2008A&ARv..15..225M}. We note, however, that evidence for the presence of a
disk does not necessarily imply that accretion powers the observed emission of
AXPs.}

{To further investigate the accretion scenario, we have taken an
alternative and purely phenomenological approach, based on the comparison of
the aperiodic variability properties of accreting objects with that of several
magnetars including the prototypical AXP \src, often suggested in literature as
an accreting system, and of two other bright magnetars, 1RXS~J170849.0$-$400910
and 1E~1841$-$045. For completeness, we also include a bright radio pulsar
PSR~B1509$-$58 for comparison.}

{If accretion is the mechanism at the base of the observed emission, one
shall expect to observe a similar aperiodic variability for accreting sources and
magnetars. Accretion is known to be an intrinsically noisy process
\citep{1997MNRAS.292..679L} and all accreting systems, from young stellar
objects to active galactic nuclei, do exhibit strong red-noise type aperiodic
variability
\citep{1971ApJ...168L..43R,1974PASJ...26..303O,2009A&A...507.1211R,2015SciA....1E0686S}. There is no reason for magnetars to be an exception. In this work, we
show, however, that the observed variability properties of AXPs are drastically
different from those of accreting systems, and thus conclude that the observed
emission is likely not powered by accretion.}

\section{Object selection, observations, and analysis}

Being the brightest and arguably the best studied AXPs, \src is also the
archetypal source discussed in the context of the accretion scenario, and, in
fact, {one of the two magnetars for which substantial} observational
arguments exist to support this interpretation {(the other being
1E~161348$-$5055 in RCW~103, \citealt{2018arXiv180305716E})}. In particular,
the detection of mid-infrared emission from a cool disk around the source
\citep{2006Natur.440..772W} largely motivated the development of the fall-back
accretion scenario.

The peculiar two-component broadband X-ray spectrum, similar to that of some
accretion-powered pulsars \citep[see i.e., Fig 1 in ][]{2012A&A...540L...1D},
has also been interpreted in favor of accretion for this source \citep[see also][and references therein]{2007ApJ...657..441E,2013ApJ...764...49T,2014ApJS..212....6O,2015MNRAS.454.3366Z,2020arXiv200407963B}. In addition, its high observed flux and
comparatively low spin period ($\sim8.7$\,s) make \src an ideal target for our
study.

{In this study, we also consider two other magnetars,
1RXS~J170849.0$-$400910 and 1E~1841$-$045, both of which are bright and have
been observed with the same instrument (i.e., \textit{NuSTAR}), which is also
used in this work for \src and the reference accreting sources. Finally, for
completeness, we also include a bright radio pulsar PSR B1509$-$58, which is also
detected in the hard X-ray band and was observed with \textit{NuSTAR}
\citep{2016ApJ...817...93C}.} {On the other hand, we have not included
1E~161348$-$5055 in our sample, since its very long spin period ($\sim6.7$
hours) would require very long observations to probe variability on timescales
comparable and exceeding the spin period.}

{Indeed,} the power spectral density (PSD) of magnetized NSs and WDs
{used to describe their variability} reveals the presence of strong
power-law noise, which is truncated around the spin frequency of the compact object
\citep{2009A&A...507.1211R}. Here, the break is associated with the interaction
of the accretion flow with the magnetosphere, which suppresses noise at higher
frequencies \citep{2009A&A...507.1211R,2019MNRAS.482.3622S}. If the origin of
the observed X-ray emission is the same in accreting pulsars and AXPs, one
would expect similar power spectra, 
that is to say red-type noise with a break
around the spin frequency. Considering that the physics of accretion is expected to
be the same for similar luminosities, the amplitude of noise relative to the
pulsed signal can also be expected to be similar for X-ray pulsars and
magnetars. We observe that the lack of secular spin-up trends of magnetar
candidates would imply, in the context of the fossil-disk accretion scenario, that
they are close to co-rotation with the inner accretion disk regions. This in
turn means that the break in the power spectrum can only occur around the spin
frequency; additionally, in the accretion scenario, the noise at lower frequencies shall
not be suppressed and must be observable. {Constraining noise level on
timescales of hours is, however, challenging from an observational perspective as
it requires very long observations, hence the omission of 1E~161348$-$5055.}

{The selection of the reference accreting sources for comparison can be
fairly arbitrary as the observed power spectra are qualitatively similar for
all accretors
\citep{2009A&A...507.1211R,2015SciA....1E0686S,2019MNRAS.482.3622S}.
Nevertheless, for a quantitative comparison, we have chosen objects which are
as similar as possible to the considered magnetars in terms of phenomenology
and the quality of the existing observations. As a reference accreting pulsar, we
have selected 1A~0535$+$262 observed with \textit{NuSTAR} in quiescence in
\cite{2019MNRAS.487L..30T}. In this observation, the source was found at a
luminosity comparable with that of AXPs and it exhibited a two-component energy
spectrum similar to that of X~Persei and \src \citep{2019MNRAS.487L..30T}. This
is an important point as the hypothesis is such that the persistent emission of AXPs is
at least, in part, based on the similarity of their spectra to that of
accretion-powered pulsars \citep{2013ApJ...764...49T}.}

{Accretion onto a white dwarf has also been suggested to power
the persistent emission of AXPs \citep{2014PASJ...66...14C,2020arXiv200407963B}. As
a reference white dwarf accretor, we have chosen the intermediate polar (IP)
GK~Persei, which was observed by \textit{NuSTAR} in outburst and thus has one of
the highest luminosities among all IPs \citep{2019MNRAS.482.3622S}. We
emphasize, however, that power spectra of all IPs appear to be qualitatively
similar \citep{2019MNRAS.482.3622S}. We stress again that high quality data
obtained with the same instrument (\textit{NuSTAR}) at comparable flux levels
are available for all considered objects.}
\begin{table*}[!ht]
        \begin{center}
        \begin{tabular}{lllllll}
                Source & \src & 1A 0535+262 & GK Per & 1RXS~J170849.0 &  1E~1841$-$045 & PSR B1509-58\\
                \hline
                obs.id & 30001023003 & 90401370001 & 30101021002 & 30401023002 & 30001025012 & 40024001002\\
                $L_{\rm x}, 10^{34}$\,erg\,s$^{-1}$ & 38.3$^{a}$ & 6.7$^{b}$  & 0.14$^c$ & 4.2$^{d}$ & 0.85$^{d}$ & $\sim40^{e}$\\
                exposure, ks & 143 & 55 & 72 & 93 & 100 & 34\\
                count-rate, s$^{-1}$ & 3.46 & 2.54 & 1.7 & 1.3 & 1.34 & 1.3\\
                $f_{\rm break}$, Hz & 0.115  & $9.662\times10^{-3}$ &  $2.849\times10^{-3}$ & 9.08676$\times10^{-2}$&  8.4825$\times10^{-2}$ & 6.59 \\
                $\Gamma_1/\Gamma_2$ & 0.7/2.0& 0.66(6)/1.89(6) & 0.72(6)/2.30(7) & 0.7/2.0 & 0.7/2.0 & 0.7/2.0\\
                $A_{\rm noise}$ & $\le2\times10^{-3}$ & 0.44$_{-0.07}^{+0.15}$& 0.37$_{-0.01}^{+0.18}$ & 2.7$_{-1.8}^{+1.8}\times10^{-3}$ & $\le3\times10^{-3}$ & $\le4\times10^{-4}$\\
                $A_{\rm noise}/A_{\rm pulse}$ & $\le0.35$ & 40$_{-14}^{+22}$ & 220$_{-125}^{+730}$ & $\le0.2$ & $\le0.17$ & $\le6\time10^{-4}$\\
        \end{tabular}
        \end{center}
        \caption{Summary of the data used and derived PSD parameters for \src, 1A~0535+262, and GK~Per. $^a$-using fluxes reported in \cite{2015ApJ...815...15W} and distance from \cite{2020arXiv200407963B}, $^b$-\cite{2019MNRAS.487L..30T}, $^c$-\cite{2019MNRAS.482.3622S}, $^{d}$-\cite{2014ApJS..212....6O}, $^e$-\cite{2016ApJ...817...93C}. All uncertainties are reported at the $1\sigma$ confidence level.}
        \label{tab:psd_fit}
\end{table*}

\subsection{Observed variability in accretors and magnetars} {As already
mentioned, all sources in the sample have been observed by \textit{NuSTAR}. The
summary of the observations used in the analysis is presented in
Table~\ref{tab:psd_fit}. To investigate the observed variability properties in
all objects, we reduced the data and} extracted {source} light curves in
the 3-80\,keV energy range, with a time resolution of 0.0625\,s using the
\texttt{HEADAS 6.27.1} software and current set of calibration files (version
20200526). In each case, the source photons were extracted from a region
centered on the source with a radius of 80$^{\prime\prime}$. The source signal
dominated the count-rate ($ \ge95$\% of all counts in all cases) so the
background was not subtracted for a timing analysis. Light curves that were extracted from
the two \textit{NuSTAR} units were corrected to the solar system barycenter and
co-added to improve the counting statistics. PSDs were constructed using the
\texttt{powspec} task and converted to a format that is readable by \texttt{XSPEC}, as
described in \cite{2012MNRAS.419.2369I}. They were also rebinned by a constant
factor to ensure that at least 20 points contribute to each frequency bin to
reduce the statistical bias \citep{10.2307/2241759} associated with the Whittle
statistics \citep{1953ArM.....2..423W} used to fit the resulting PSDs.

\begin{figure}[!t]
        \centering
                \includegraphics[width=\columnwidth]{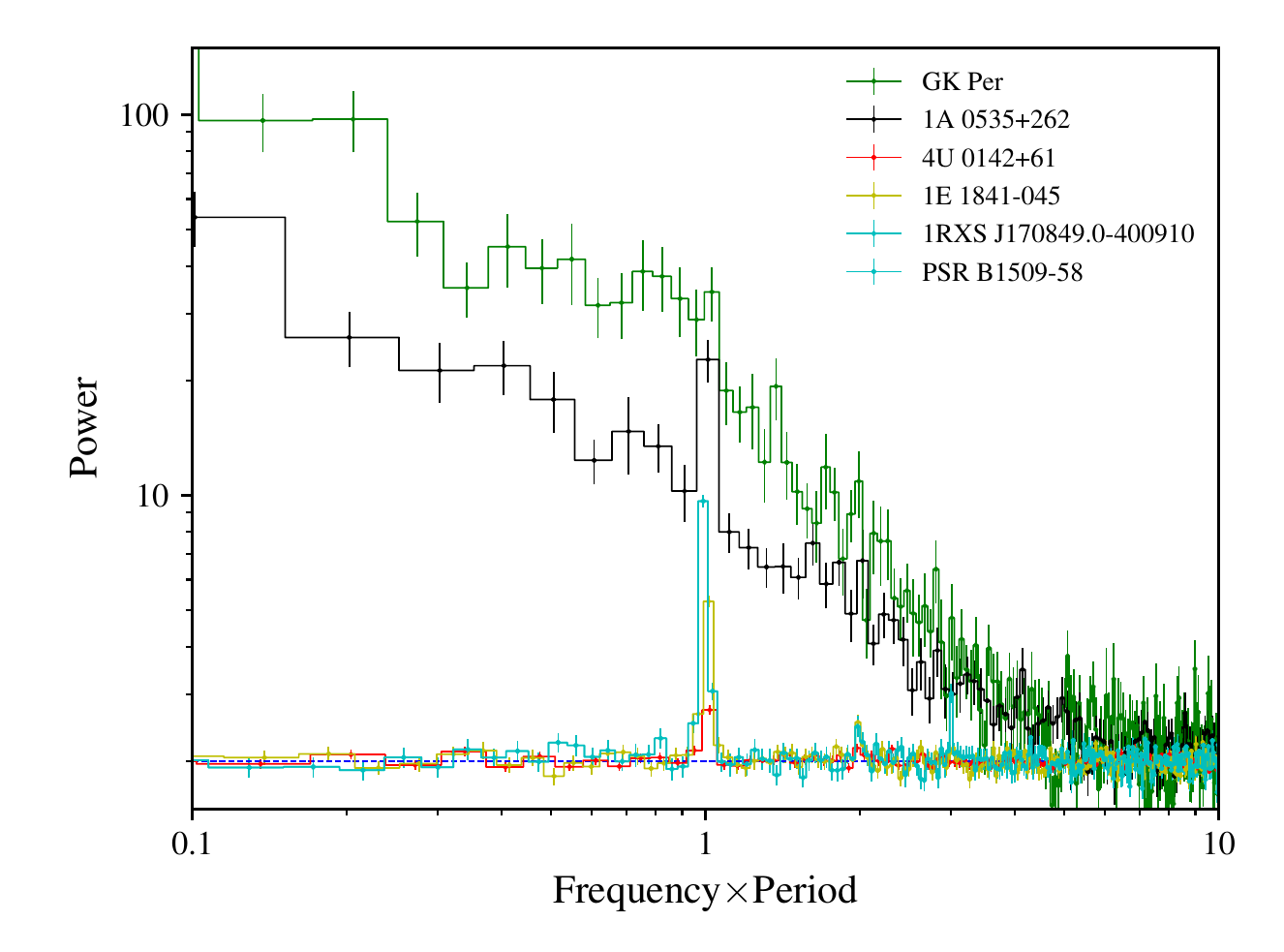}
        \caption{Power density spectra of the magnetized compact objects discussed in the text and labeled in the legend. The frequency is expressed in units of respective objects spin frequency and power is normalized such that expected white noise level is 2.}
        \label{fig:psd}
\end{figure}

To model the PSDs, we chose a broken power law with a break fixed at the spin
frequency of a given source (see i.e., Table~\ref{tab:psd_fit}) because all
three sources are expected to be close to co-rotation. A zero-width Lorentzian
curve with the same frequency and an additional constant were also included in
the model to account for pulsations and white noise. The white noise amplitude
was fixed to the expected level of two, but it was not subtracted for a clearer
presentation of the power spectrum of the AXPs. We also rescaled the frequency
axis for plotting so that the break in the PSDs appears at the same location to
ease the comparison between individual objects, as was done by
\cite{2009A&A...507.1211R}. The results are shown in Fig.~\ref{fig:psd}.

As it is clearly seen from the figure, low-frequency noise dominates the power
spectra of the two, well-established accreting objects, 1A~0535+262 and GK~Per.
On the contrary, it is completely absent in the PSD of {magnetars}. To
estimate an upper limit on the noise amplitude, we included a broken power law
component, fixing the indices {$\Gamma_{1,2}$ below and above the break
frequency $f_{\rm break}$} to values similar to that obtained for the other two
reference sources, and we calculated the $1\sigma$ confidence bounds for the
amplitudes {$A_{\rm noise/pulse}$ of the noise and pulsed signal} using
the \texttt{error} command in \texttt{XSPEC}. The results are presented in
Table~\ref{tab:psd_fit}.

The noise amplitude in \src is consistent with zero and, if there is any, it has to be at
least by factor of 180 lower than that in the other two objects. In Table 1, we also report the ratio of the pulsed signal amplitude to noise
amplitude. Also, in this case, the relative noise power in \src is substantially
lower (by a factor of at least 75) than that of the reference accreting sources.
{A similar conclusion also holds for the other two considered
magnetars.} We note that, to our best knowledge, there are also no reports of
aperiodic variability for {the persistent emission in} the other magnetars
not studied here. Therefore, we conclude that the sources studied in this work
do not constitute a special sample, and the absence of any observed aperiodic
variability is a strong general argument against the accretion-powered origin
of emission from magnetars.

\section{Conclusions} 
The recent discovery of the the two-component X-ray spectra typical of AXPs in
several accreting X-ray pulsars at low luminosity
\citep[see][]{2019MNRAS.483L.144T, 2019MNRAS.487L..30T} has suggested that
{such a spectral shape} might be a common feature of {accretion-powered pulsars at low luminosities. The similarity of the spectra of these
objects and AXPs \citep{2012A&A...540L...1D} can thus be viewed as an argument
in favor of the common origin of the observed emission in X-ray pulsars and AXPs,
that is to say accretion \citep{2012A&A...540L...1D,2013ApJ...764...49T}.} We examined,
{therefore,} the hypothesis that the persistent emission of  AXP \src\ 
{could indeed be} powered by accretion.

We find, however, that, despite the similarity of the observed X-ray spectra,
the aperiodic variability that is universally observed in all accreting systems,
including the low-luminous X-ray pulsars and accreting white dwarfs
\citep{2009A&A...507.1211R,2015SciA....1E0686S}, is completely absent in \src
{RXS~J170849.0$-$400910 and 1E~1841$-$045.} More specifically, this
result follows from comparing the observed PSDs of {these objects} with
that of the accreting pulsar 1A~0535+262 (observed in the low luminosity state)
and of the intermediate polar GK~Persei, {which can be considered as
representative objects of an accreting neutron star and a white dwarf,
respectively}. All of these objects were observed with the \textit{NuSTAR}
observatory at a comparable flux and luminosity level, which allowed us to obtain
power spectra of a similar quality. We find that despite {similar
luminosities, counting statistics, and energy spectra,} the variability
properties of accreting objects and magnetars are {drastically}
different and no evidence for the low-frequency red noise that is typical for accreting
sources is detected in the magnetars of the sample. We emphasize that the
choice of other reference objects would not alter our conclusions since
aperiodic variability is an established feature of accreting systems.

{We conclude, therefore, that the observed persistent emission from
\src, 1RXS~J170849.0$-$400910, 1E~1841$-$045, and PSR~B1509$-$58 is not due to
accretion, as expected.} Considering that \src is a prime candidate for
accretion-powered AXPs and the lack of detected (or reported) variability in
any of the magnetar candidates, we conclude that our finding constitutes a
strong independent argument against the accretion-powered origin of the
persistent X-ray emission in magnetars. This conclusion can be further verified
by extending a similar analysis to more sources.

\begin{acknowledgements} VD thanks Joachim Tr\"umper for asking the question
whether accreting pulsars in low-luminosity state are indeed magnetars. A
question which this paper aims to answer. This work was supported by the
Russian Science Foundation (grant 19-12-00423). VFS thanks Deutsche
Forschungsgemeinschaft for financial support (grant DFG-GZ WE 1312/53-1). We
thank German Academic Exchange Service (DAAD, project 57405000) and the Academy
of Finland (projects 324550, 331951) for travel grants.

\end{acknowledgements}

\vspace{-0.3cm}
\bibliography{biblio}   
\vspace{-0.3cm}

\end{document}